\documentclass[12pt,letterpaper]{article}
\pdfoutput=1

\usepackage{xcolor}
\usepackage{amsfonts,amssymb,amsmath}
\usepackage{MnSymbol,wasysym}
\usepackage{esint}
\usepackage{comment}

\usepackage{graphicx}
\usepackage{amsmath,systeme}
\usepackage{caption}
\usepackage{subcaption}
\usepackage{booktabs}

\usepackage{listings}

\usepackage{hyperref}

\usepackage[style=numeric-comp, sorting=none]{biblatex}
\DeclareFieldFormat{doi}{}
\DeclareFieldFormat{url}{}
\DeclareFieldFormat{eprint}{}

\numberwithin{equation}{section}

\newcommand{\be}{\begin{equation}}
\newcommand{\ee}{\end{equation}}
\newcommand{\ben}{\begin{displaymath}}
\newcommand{\een}{\end{displaymath}}
\newcommand{\bea}{\begin{eqnarray}}
\newcommand{\eea}{\end{eqnarray}}
\newcommand{\bean}{\begin{eqnarray*}}
\newcommand{\eean}{\end{eqnarray*}}

\def\g {\gamma}

\usepackage{amsmath}

\newcommand{\beq}{\begin{equation}}
\newcommand{\eeq}{\end{equation}}
\newcommand{\beqr}{\begin{displaymath}}
\newcommand{\eeqr}{\end{displaymath}}
\newcommand{\beqa}{\begin{eqnarray}}
\newcommand{\eeqa}{\end{eqnarray}}
\newcommand{\beqar}{\begin{eqnarray*}}
\newcommand{\eeqar}{\end{eqnarray*}}
\newcommand{\cmmnt}[1]{}

\usepackage{setspace}

\usepackage{float}

\allowdisplaybreaks

\newcounter{customfootnote}  % Create a new counter
\newcommand{\customfootnote}[1]
{%
  \refstepcounter{customfootnote}%
  \footnote[\value{customfootnote}]{#1}%
}

\renewcommand{\paragraph}[1]{\noindent \textbf{#1}\quad}

\begin{document}
%%%%%%%%%%%%%%%%%%%%%%%%%%%%%%%%%%%%%%%%%%%%%%%%%%%%%%%%%%%%%%%%%%%%%%%%
%%%%%%%%%%%%%%%%%%%%%%%%%%%%%%%%%%%%%%%%%%%%%%%%%%%%%%%%%%%%%%%%%%%%%%%%
%%%%%%%%%%%%%%%%%%%%%% TITLEPAGE %%%%%%%%%%%%%%%%%%%%%%%%%%%%%%%%%%%%%%%
%%%%%%%%%%%%%%%%%%%%%%%%%%%%%%%%%%%%%%%%%%%%%%%%%%%%%%%%%%%%%%%%%%%%%%%%
%%%%%%%%%%%%%%%%%%%%%%%%%%%%%%%%%%%%%%%%%%%%%%%%%%%%%%%%%%%%%%%%%%%%%%%%
%\onehalfspacing

%\title{\LARGE \bf Stationary and Self Similar solutions of Gross-Pitaevskii Equation in 2D and 3D, and their Borel analysis}
\title{\LARGE \bf Acoustic black holes, white holes, and wormholes in Bose-Einstein condensates in two dimensions}

\author{
	 Sachin Vaidya$^1$,
	 Martin Kruczenski$^1$\thanks{E-mail: \texttt{vaidya2@purdue.edu, markru@purdue.edu.}} \\
	$^1$ Dep. of Physics and Astronomy, \\ Purdue University, W. Lafayette, IN  \\
	}

\date{}

\maketitle

\begin{abstract}
In a previous article, we studied stationary solutions to the dynamics of a Bose-Einstein condensate (BEC) corresponding to acoustic (or Unruh) black/white holes, namely configurations where the flow becomes supersonic creating a horizon for phonons.  In this paper, we consider again the Gross-Pitaevskii Equation (GPE) but looking for stationary numerical solutions in the case where the couplings are position dependent in a prescribed manner. Initially we consider a 2D quantum gas in a funnel-like spatial metric. We then reinterpret this solution as a solution in a flat metric but with spatially dependent coupling and external potential. In these solutions the local speed of sound and magnitude of flow velocity cross, indicating the existence of a supersonic region and therefore of sonic analogues of black/white holes and wormholes. We discuss the numerical techniques used. We also study phase (and density) fluctuations in these solutions and derive approximate acoustic metric tensors. For certain external potentials, we find uniform density acoustic black hole configurations and obtain their Hawking temperature.
\end{abstract}

\clearpage
\newpage

%\makeindex
\tableofcontents
%\keywords{Classical string solutions, AdS/CFT, Wilson loops}

%\preprint{\tt{} \\
%          \tt{hep-th/yymmnnn}  }

\onehalfspacing
%%%% INTRODUCTION
%\setcounter{section}{1}
\section{Introduction}
\label{intro}
%\section{Gross Pitaevskii Equation}
In \cite{PhysRevLett.46.1351} Unruh proposed the idea of studying acoustic black holes which can help better understand quantum phenomena such as Hawking radiation \cite{HawkingS.W.1975Pcbb} from the acoustic analog. Such analog gravity models have been theoretically studied extensively \cite{Barceló2011, CarlosBarceló_2001, Visser2002, PhysRevLett.85.4643, PhysRevD.105.124066, Tian2022} in the past. Such models have also been studied in cosmological context (see \cite{PhysRevA.69.033602}, \cite{PhysRevLett.91.240407}). In a previous paper\cite{vaidya2024stationaryacousticblackhole}, we studied stationary singular acoustic black hole configurations (in 2D) numerically and through series expansions by finding stationary solutions of the Gross-Pitaevskii equation in 2D with circular symmetry. Asymptotically at infinity the fluid is at rest, but as we approach the origin there is a radially inward flow that becomes supersonic at a certain radius. At that radius, we therefore have an acoustic horizon. Sound cannot escape the supersonic region. Since the fluid accumulates near the origin, we have a singularity in the density that in experiments will be resolved, for example,  by three-body recombination (see \cite{PhysRevA.109.023305}, \cite{Tamura:2023mby}) or other effects that remove atoms from the condensate. Those solutions were for uniform coupling and no external potential. However, experimental techniques allow for more general cases with external potentials and position-dependent coupling. We can also conceive that future developments allow us to confine the gas to surfaces with non-trivial curvature allowing for an external non-trivial spatial metric.  
In the general case with spatial metric $\g_{ij}(\Vec{\mathbf{r}})$, coupling $g_0(\Vec{\mathbf{r}})$, and external potential $V(\Vec{\mathbf{r}})$, 
the Gross-Pitaevskii equation (GPE) \cite{NozieresPinesBEC} for $\psi(\Vec{\mathbf{r}},t)$ reads 
\begin{equation} \label{e:GPE} 
%\begin{align}
i \hbar  \frac{\partial \psi}{\partial t} = -\frac{\hbar^{2}}{2 m} \frac{1}{\sqrt{\g}} \partial_i \left(\sqrt{\g} \g^{ij}\partial_j \psi\right) + g_0(\Vec{\mathbf{r}})\, {\lvert \psi  \rvert}^{2}\, \psi +V(\Vec{\mathbf{r}}) \psi  
%\end{align}
\end{equation}
If the parameters of the equation and its solution are slowly varying, we can define a local speed of sound \cite{NozieresPinesSound}
\begin{equation} \label{e:speed of sound}
c\left(\Vec{\mathbf{r}} , t \right)=\sqrt{\frac{n\left(\Vec{\mathbf{r}} , t \right)g_0(\Vec{\mathbf{r}})}{m}}
\end{equation}
where,  $n=|\psi|^2$ is the density. We can also define a radial flow velocity given by \cite{NozieresPinesVelocity}
\begin{equation} \label{e:flow velocity}
\Vec{v} \left(\Vec{\mathbf{r}} , t \right) = \frac{\Vec{\mathbf{j}} \left(\Vec{\mathbf{r}} , t\right)}{n \left(\Vec{\mathbf{r}} , t\right)} = \frac{\hbar}{m} \partial^{r} \theta  \left(\Vec{\mathbf{r}} , t\right) \hat{\mathbf{r}}
\end{equation}

\section{Stationary states: Acoustic black holes, white holes, and wormholes} \label{sec:general}
To get the stationary solutions, we first substitute 
\beq
\psi  \left(\Vec{\mathbf{r}} , t\right) = {\mathrm e}^{-i \mu  t} \phi  (\Vec{\mathbf{r}},t)
\eeq
into the Gross-Pitaevskii equation (GPE) \eqref{e:GPE}, and for the purposes of calculation, we scale all equations with
\beqa \label{e:scaling}
T&=&\mu t, \nonumber\\
\Vec{\mathbf{R}}&=&\sqrt{\frac{2m\mu}{\hbar}}\Vec{\mathbf{r}}, \nonumber\\
\Phi &=& \frac{\phi}{\sqrt{\hbar\mu}}, \nonumber\\
\mathcal{V} &=& \frac{V}{\hbar\mu}
\eeqa
which turns the Gross-Pitaevskii equation into
\begin{equation} \label{e:GPE2Scaled}
%\begin{align}
\Phi  + i \frac{\partial \Phi  }{\partial T} = - \frac{1}{\sqrt{\g}} \partial_i \left(\sqrt{\g} \g^{ij}\partial_j \Phi\right)+g_0(\Vec{\mathbf{r}})\, {\vert \Phi  \rvert}^{2} \Phi + \mathcal{V}(\Vec{\mathbf{r}})\, \Phi
%\end{align}
\end{equation}
In this work, we consider quantum gases in 2D and use a conformally flat background metric\customfootnote{In 2D any metric can be put in conformally flat form with an appropriate change of coordinates}{}:
\beq
ds^2 = f(R) \left[dR^2 + R^2 d\varphi^2\right] 
\eeq
which can also be written as
\beqa \label{e:funnel_meric}
\g_{i j} = f(R)
\left[\begin{array}{ccc}
1 & 0 
\\
0 &  R^2
\end{array}\right]
\eeqa
The GP \eqref{e:GPE2Scaled} equation in this non-trivial background becomes
\begin{equation} \label{e:GPE2ScaledB}
%\begin{align}
\Phi  + i \frac{\partial \Phi  }{\partial T} = - \frac{1}{f(R)} {\nabla_{R}}^{2}\Phi +g_0(\Vec{\mathbf{R}})\, {\vert \Phi  \rvert}^{2} \Phi + \mathcal{V}(\Vec{\mathbf{R}})\, \Phi
%\end{align}
\end{equation}
where, ${\nabla_{R}}^2$ is the usual Laplacian  and $\Phi=\mathrm{\Phi(T,R,\varphi)}$ is the scaled wave function. Further, assuming only radial dependence and replacing 
\beq \label{e:substitute}
\Phi  (R) = \rho  (R) \exp(i \theta(R)))
\eeq
The imaginary part of the equation gives
\beq \label{e:phase_grad}
 \frac{\partial\theta}{\partial R} = \frac{B}{R\,\rho^2}
\eeq
where $B$ is a constant of integration. Finally, after the substitution \eqref{e:substitute} the real part of \eqref{e:GPE2ScaledB} using \eqref{e:phase_grad} reads
\begin{equation} \label{e:StationaryODEmetricA}
%\begin{align}
\frac{d^{2} \rho }{d R^{2}}+\frac{1}{R} \frac{d \rho }{d R}-\frac{B^{2}}{\rho^{3} R^{2}}+f(R) \left[(1-\mathcal{V}(R))\rho-g_0(R)\rho^{3} \right]  = 0
%\end{align}
\end{equation}
This is the most general form of the static equation that we consider. 

\paragraph{Correspondence principle of solutions:}
Since the form of equation \eqref{e:StationaryODEmetricA} depends only on three functions, if we consider two sets of external metrics $f_{1,2}(R)$, potentials $\mathcal{V}_{1,2}(R)$ and couplings $g_{01,02}(R)$ such that
\beqa
 f_1(R) (1-\mathcal{V}_1(R))  &=&  f_2(R) (1-\mathcal{V}_2(R))  \\
 f_1(R)\, g_{01}(R)  &=& f_2(R)\, g_{02}(R) 
\eeqa
then these two sets of functions have the exact same set of solutions for $\rho(R)$. As we see later, unfortunately, this correspondence does not extend to time-dependent fluctuations around these stationary background solutions. 

 Going back to \eqref{e:StationaryODEmetricA}, since this equation cannot be solved analytically, we solve it using the same boundary value problem (BVP) techniques (Newton iteration with Chebyshev collocation and Physics-Informed Neural Network) we used in \cite{vaidya2024stationaryacousticblackhole} for singular stationary solutions. We also study simpler versions of these solutions in the form of uniform density solutions by introducing a different external potential. Then we study the dynamics of fluctuations in the density and the phase of these solutions under some approximations.

\section{Acoustic wormhole solution} \label{sec:wormhole BVP}

In this section we consider a case with no external potential and uniform coupling that can be taken as $g_0(R)=1$ by a rescaling. In this case we can introduce a scaled local speed of sound of the form 
\beq \label{e:scaled_sound_funnel}
C=\frac{\sqrt{2}\rho\left(R \right)}{\sqrt{f\left(R \right)}}
\eeq
and a scaled flow velocity (from equation \eqref{e:phase_grad})
\beq \label{e:scaled_vel_funnel}
\Vec{\mathbf{V}}  \left(R \right) = V^R \left(R \right) \hat{\mathbf{R}} = \frac{2 \partial_R \theta \left(R \right)}{f(R)} \hat{\mathbf{R}} = \frac{2 B}{f\left(R \right) R (\rho\left(R \right))^2} \hat{\mathbf{R}}
\eeq
From \eqref{e:scaled_vel_funnel} it is clear that $B<0$ corresponds to inward flow and $B>0$ corresponds to outward flow, and both have the same flow speed for a given $\lvert B\rvert$. 

\begin{figure} 
\centering
    \includegraphics[width=0.6\columnwidth]{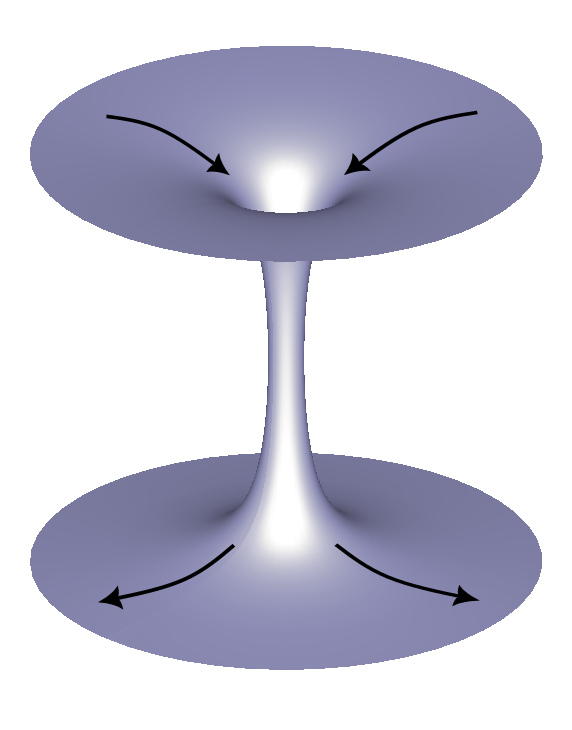}
    %\vspace{5pt}
    \caption{Theoretical wormhole configuration. The fluid moves along a non-trivial surface connecting two asymptotic regions. The motion is supersonic in the vertical part. In principle the outer regions can be reconnected to recirculate the subsonic fluid without requiring sources or sinks from experimental point of view.} \label{fig:Wormhole configuration}
\end{figure}

Now we consider 
\beq\label{fR}
f(R) = 1+\left(\frac{R_0}{R}\right)^4
\eeq
along with $\mathcal{V}(R)=0$ and $g_0(R)=1$ in \eqref{e:StationaryODEmetricA}, giving
\begin{equation} \label{e:StationaryODEmetric}
%\begin{align}
\frac{d^{2} \rho }{d R^{2}}+\frac{1}{R} \frac{d \rho  }{d R}+(\rho -\rho  ^{3}) \left(1+\left(\frac{R_0}{R}\right)^4\right)-\frac{B^{2}}{\rho^{3} R^{2}} = 0
%\end{align}
\end{equation}
The background spatial metric $ds^2 = \left(1+\left(\frac{R_0}{R}\right)^4\right) (dR^2 + R^2 d\varphi^2)$ transforms to $ds^2 = \left(1+\left(\frac{R_0}{u}\right)^4\right) (du^2 + u^2 d\varphi^2)$ under $u=\frac{{R_0}^2}{R}$ which indicates a symmetry about $R_0$. Equation \eqref{e:StationaryODEmetric} also clearly exhibits this symmetry and therefore has a similar behavior at $R=0$ as $R\to\infty$. We look for solutions of the form $\rho(R)\to 1$ as $R\to\infty$, and by symmetry  $\rho(R)\to 1$ as $R\to 0$. This metric then describes two separate asymptotic flat regions (see appendix \ref{Appendix A})
%\customfootnote{It is obvious from a better coordinate transformation such as $\ln\left(\frac{R}{R_{0}}\right) = u$}{} 
where the fluid flows from one to the other. For that reason, we call this configuration an acoustic wormhole as illustrated in figure \ref{fig:Wormhole configuration}. 

\subsection{Numerical solutions}
Equation \eqref{e:StationaryODEmetric} cannot be solved analytically. Therefore, solutions are obtained by using Newton iteration with Chebyshev collocation (with iteration over $R_{0}$) as well as using deep learning, as discussed in the case of singular stationary solutions in \cite{vaidya2024stationaryacousticblackhole}.

\paragraph{Newton iteration with Chebyshev collocation:} Here we have a symmetry about $R_{0}$ and therefore we need to solve only for $R_{0}\leq R < \infty$ numerically by mapping it to $[-1,1]$ as follows:
\beq
R = R_0 + A \frac{1+U}{1-U}
\eeq
where we choose $A=10$ to better resolve the structure of the solution. For Chebyshev collocation, we use the collocation grid points given by $U_k = \cos\left(\frac{\pi k}{N} \right)$ where $k=0,1,2,\ldots,N$. The derivatives are discretized on these grid points using Chebyshev collocation differential matrices (see \cite{SALEHTAHER20134634}). 

\begin{figure} 
\centering
    \includegraphics[width=\linewidth]{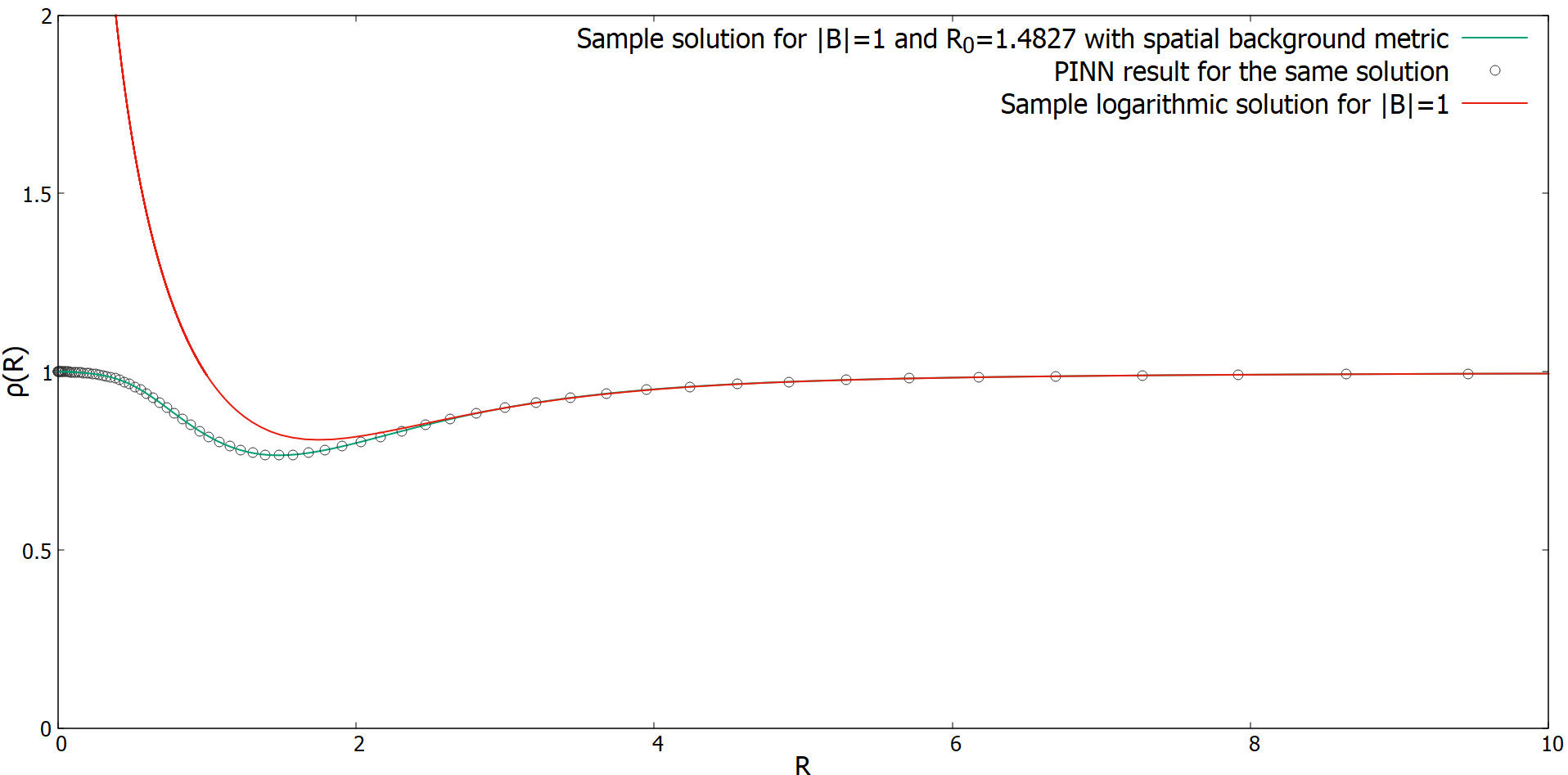}
    %\vspace{5pt}
    \caption{Sample solution with background metric compared with logarithmic singular solution for $|B|=1$ in 2D. Neural Network result for the same solution shown in circles.} \label{fig:sample solution with metric}
    \bigskip
    \includegraphics[width=\linewidth]{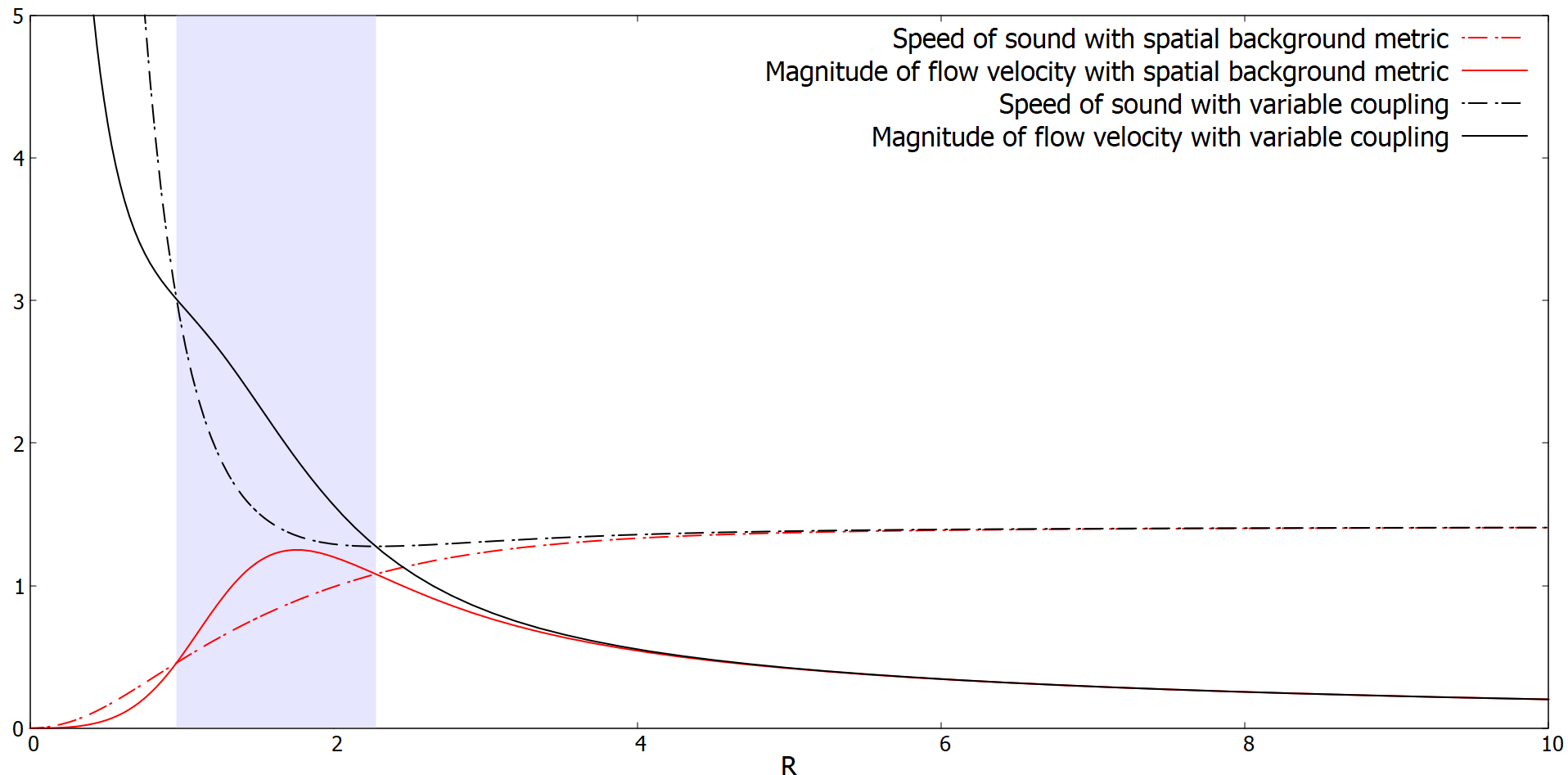}
    %\vspace{5pt}
    \caption{Speed of sound (dashed line) and magnitude of flow velocity (solid line) for $\lvert B \rvert=1$ and $R_0=1.4827$ with background metric (in red) and variable coupling (in black) in 2D with supersonic region in blue shade.} \label{fig:c and v with metric}
\end{figure}

To construct a Newton method \cite{GALANTAI200025} for a boundary value problem, we first write the differential equation \eqref{e:StationaryODEmetric} in variable $U$ as
\begin{align}\label{e:StationaryODE=G=0}
ODE & = G  \left(\rho  \left(U \right),  \rho'  \left(U \right), \rho''  \left(U \right)\right) = 0 \nonumber \\ 
    & \text{ where}, \rho'  \left(U \right) = \frac{d \rho  \left(U \right)}{d U}, \rho''  \left(U \right) = \frac{d^{2} \rho  \left(U \right)}{d U^{2}} 
\end{align}
and we expand $G$ in~\eqref{e:StationaryODE=G=0} about $\rho_\text{old}(U)$ at first order  \cite{na1979computational}
\begin{align} \label{e:StationaryODETaylorexpandG}
\lefteqn{G_\text{old} + (\rho_\text{new} - \rho_\text{old}) \left(\left.\frac{d G}{d \rho}\right\vert_{\rho=\rho_\text{old}} \right)}\nonumber \\ & \indent\indent + (\rho_\text{new}' - \rho_\text{old}') \left(\left.\frac{d G}{d \rho'}\right\vert_{\rho'=\rho_\text{old}'} \right)+ (\rho_\text{new}'' - \rho_\text{old}'') \left(\left.\frac{d G}{d \rho''}\right\vert_{\rho''=\rho_\text{old}''} \right) = 0
%\addtocounter{equation}{1}\tag*{\normalsize(\theequation)}
\end{align}
We use \eqref{e:StationaryODETaylorexpandG} along with Chebyshev collocation differential matrices to implement the Newton iteration (in C++) with mixed boundary conditions $\rho'(U=-1)=0$ and $\rho(U=1)=1$. This can be easily achieved by adding a row for the derivative at the left boundary (at $R_0$) in the rectangular matrix of the system of $N-1$ discretized equations in $N$ unknowns. Then these Newton iterations are performed with some initial guess of the function as well as $R_{0}$ until the iteration error becomes small enough and falls within the acceptable numerical error. It is interesting to note that the solutions obtained using this approach turn out to have a lower bound on the value of $R_{0}$ for a given value of the parameter\customfootnote{$B<0$ and $B>0$ have the same solutions for a given $\lvert B\rvert$.} $\lvert B\rvert$. A sample solution is shown in figure \ref{fig:sample solution with metric} and plotted against a corresponding singular stationary solution from \cite{vaidya2024stationaryacousticblackhole} to show how this  model provides a regularization of the singular solutions discussed in \cite{vaidya2024stationaryacousticblackhole}.

\paragraph{Physics-Informed Neural Network:} This problem can also be solved using neural networks. We employ the same neural network\customfootnote{1, 15, 20, 15, 1 neurons in each layer from input to output layers, with Adam optimizer, and $\tanh$ activation function, as well as Python library PyTorch along with automatic differentiation engine called autograd.} we used in \cite{vaidya2024stationaryacousticblackhole} except that now we map $R\in[0,\infty]\to U\in[-1,1]$ with $R=R_{0}\frac{1+U}{1-U}$. We scale the unknown function as $1 - 0.1\left(1-U^2\right)^{2} N(U, P)$ so that the only unknown is the neural network $N(U, P)$ and the function and its slope are always $1$ and $0$, respectively, at both boundaries. The result of this for the same solution obtained earlier with the Newton method is shown in circles in Figure \ref{fig:sample solution with metric}. 

Figure \ref{fig:c and v with metric} shows the speed of sound and the magnitude of the flow velocity for this solution with a background funnel metric clearly indicating a crossover between the speed of sound and the magnitude of flow velocity and thus the existence of a supersonic region and, therefore, a sonic black/white hole. Since equation \eqref{e:StationaryODEmetric} and by extension its solution exhibit symmetry about $R_{0}$, regions $R\to\infty$ and $R\to 0$ represent two identical asymptotically uniform density regions. Therefore, this black/white hole can be thought of as an acoustic (one-way) wormhole (see Fig. \ref{fig:Wormhole configuration}), as we shall see.

\subsection{Fluctuations and acoustic wormhole metric}
Fluctuations of these solutions satisfy \eqref{e:GPE2ScaledB} with $\mathcal{V}(R)=0$ and $g_0(R)=1$.
Introducing density and phase perturbations of the form
\begin{align} \label{e:den phase pert}
\Phi (R,\varphi,T) & = \sqrt{n(R,\varphi,T)} \exp(i \theta(R,\varphi,T)) \nonumber \\ 
& = \sqrt{n_{0} \! \left(R \right) + n_{1} \! \left(R , \varphi , T\right)} \exp{(i (\theta_{0} \! \left(R \right) + \theta_{1} \! \left(R , \varphi , T\right)))} \nonumber \\ 
& \approx \sqrt{n_{0} \! \left(R \right)} \exp(i \theta_{0} \! \left(R \right)) \left(1 + \frac{n_{1} \! \left(R , \varphi , T\right)}{2 n_{0} \! \left(R \right)} + i \theta_{1} \! \left(R , \varphi , T\right)\right) \nonumber \\
& = \sqrt{n_{0} \! \left(R \right)} \exp(i \theta_{0} \! \left(R \right)) \left(1 + N_1 \left(R , \varphi , T\right) + i \theta_{1} \! \left(R , \varphi , T\right) \right)
\end{align}
where $n_{0} \! \left(R \right) = {(\rho_{0} \! \left(R \right))}^2$ and $\theta_{0} \! \left(R \right)$ are the background density and phase,  we get the following linearized coupled equations that govern the dynamics of density and phase perturbations. 
\begin{align} \label{e:funnel_density_pert}
%\begin{align}
\lefteqn{\left({\partial}_T - \frac{2 \lvert B\rvert}{f\left(R \right) R (\rho_{0} \left(R \right))^2} {\partial}_R \right) N_1 \left(R , \varphi , T\right)=} \nonumber\\ &\indent\indent - \frac{1}{f(R) {{\rho}_0 (R)}^2} {\nabla}_R \cdot ({{\rho}_0 (R)}^2 {\nabla}_R {\theta}_1 \left(R , \varphi , T\right))
%\end{align}
%\addtocounter{equation}{1}\tag*{\normalsize(\theequation)}
\end{align}

\begin{align} \label{e:funnel_phase_pert}
%\begin{flalign}
\lefteqn{\left({\partial}_T - \frac{2 \lvert B\rvert}{f\left(R \right) R (\rho_{0} \left(R \right))^2} {\partial}_R \right) {\theta}_1 \left(R , \varphi , T\right)=} \nonumber\\ &\indent\indent \frac{1}{f(R) {{\rho}_0 (R)}^2} {\nabla}_R \cdot ({{\rho}_0 (R)}^2 {\nabla}_R N_1 \left(R , \varphi , T\right)) \nonumber \\ &\indent\indent\indent - 2 {{\rho}_0 (R)}^2 N_1 \left(R , \varphi , T\right)
%\end{flalign}
%\addtocounter{equation}{1}\tag*{\normalsize(\theequation)}
\end{align}
where, $N_1 \left(R , \varphi , T\right) = \frac{n_{1} \! \left(R , \varphi , T\right)}{2 n_{0} \! \left(R \right)} = \frac{n_{1} \! \left(R , \varphi , T\right)}{2 \left(\rho_{0} \! \left(R \right) \right)^2}$

Under the hydrodynamic approximation ${\nabla}_R \cdot ({{\rho}_0 (R)}^2 {\nabla}_R N_1 \left(R , \varphi , T\right)) \approx 0$ (see \cite{CarlosBarceló_2001} and \cite{Visser2002}), equations \eqref{e:funnel_density_pert} and \eqref{e:funnel_phase_pert} can be combined, and the result looks like a massless scalar field $\theta_{1} \left(R , \varphi , T\right)$ in an acoustic (space-time) metric background (see \cite{PhysRevA.63.023611}). The acoustic metric here looks like the Schwarzschild metric in Painlev\'e-Gullstrand coordinates (see \cite{HamiltonAndrewJ.S.2008Trmo}).

\begin{flalign} \label{e:acoustic metric tensor}
g_{\mu \nu} & \sim \rho_{0} \! \left(R \right)^{4}
\left[\begin{array}{ccc}
-\left(1-\frac{2 B^{2} R^{2}}{\rho_{0} \! \left(R \right)^{6} \left(R^{4}+R_{0}^{4}\right)}\right) & \frac{{| B |}}{R \rho_{0} \left(R \right)^{4}} & 0 
\\
 \frac{{| B |}}{R \rho_{0} \left(R \right)^{4}} & \frac{R^{4}+R_{0}^{4}}{2 R^{4} \rho_{0} \left(R \right)^{2}} & 0 
\\
 0 & 0 & \frac{R^{4}+R_{0}^{4}}{2 R^{2} \rho_{0} \left(R \right)^{2}} 
\end{array}\right]
\end{flalign}

\begin{comment}
Under the coordinate transformation $\ln\left(\frac{R}{R_0} \right) = u$, the metric \eqref{e:acoustic metric tensor} becomes,
\begin{equation} \label{e:acoustic metric tensor wormhole}
g_{\mu \nu} \sim \rho_{0} \! \left(u \right)^{4}
\left[\begin{array}{ccc}
-\left(1-\frac{B^{2}}{R_{0}^{2} \rho_{0} \left(u \right)^{6} \cosh \left(2 u \right)}\right) & \frac{{| B |}}{\rho_{0} \left(u \right)^{4}} & 0 
\\
 \frac{{| B |}}{\rho_{0} \left(u \right)^{4}} & \frac{R_{0}^{2} \cosh \left(2 u \right)}{\rho_{0} \left(u \right)^{2}} & 0 
\\
 0 & 0 & \frac{R_{0}^{2} \cosh \left(2 u \right)}{\rho_{0} \left(u \right)^{2}} 
\end{array}\right]
\end{equation}
which is asymptotically ($u\rightarrow\infty$ and $u\rightarrow-\infty$) flat and is connected at $u=0$. In these coordinates, the condensate flows with a subsonic flow from $u\rightarrow\infty$ to $u\rightarrow-\infty$ with the flow becoming supersonic in some neighborhood of $u=0$. In coordinate $u$, the connected black hole ($R > R_0$) and white hole ($R < R_0$) look like a (one-way) wormhole (see Fig. \ref{fig:Wormhole configuration}) connecting two patches of this analog space-time\customfootnote{We consider solutions with $B<0$. It is a one-way wormhole because when there are acoustic horizons, phonons can cross the horizons only in one direction.}. 
\end{comment}
\noindent Here, condensate flows with a subsonic flow from an asymptotically flat region at $R\rightarrow\infty$ to an identical asymptotically flat region at $R\rightarrow 0$ (see appendix \ref{Appendix A}) with the flow becoming supersonic in some neighborhood of $R_0$. This looks like a (one-way) wormhole (see Fig. \ref{fig:Wormhole configuration}) connecting two patches of this analog space-time\customfootnote{We consider solutions with $B<0$. It is a one-way wormhole because when there are acoustic horizons, phonons can cross the horizons only in one direction.}.

\section{Uniform density wormholes}
\begin{figure}
    \centering
    \begin{subfigure}{1\textwidth}
        \centering
        \includegraphics[width=\linewidth]{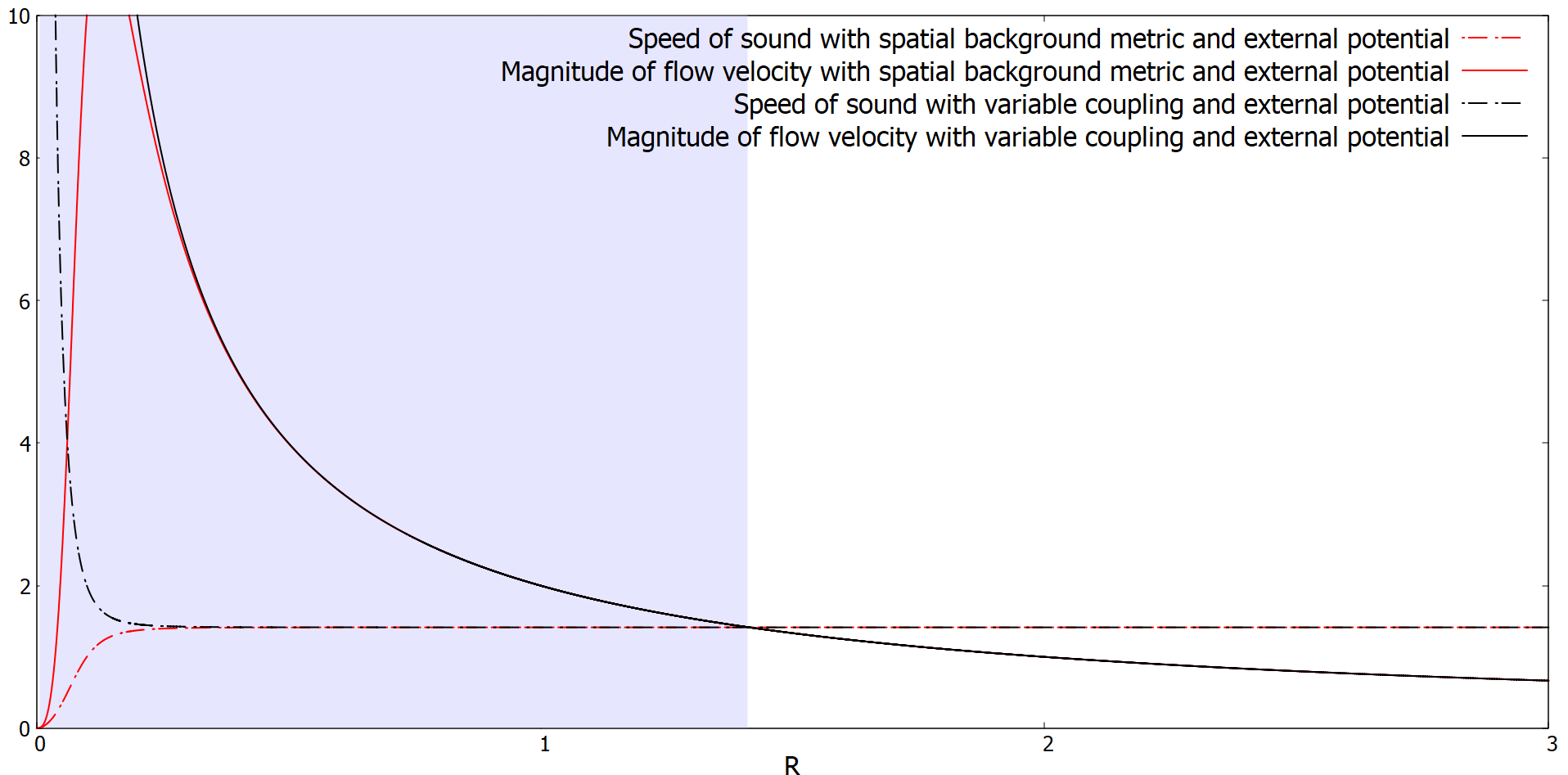}
        \caption{$|B|=1.0$ and $R_0=0.1$} \label{fig:B_1.0_r0_0.1_pot}
    \end{subfigure}
    \\[10pt]
    \begin{subfigure}{1\textwidth}
        \centering
        \includegraphics[width=\linewidth]{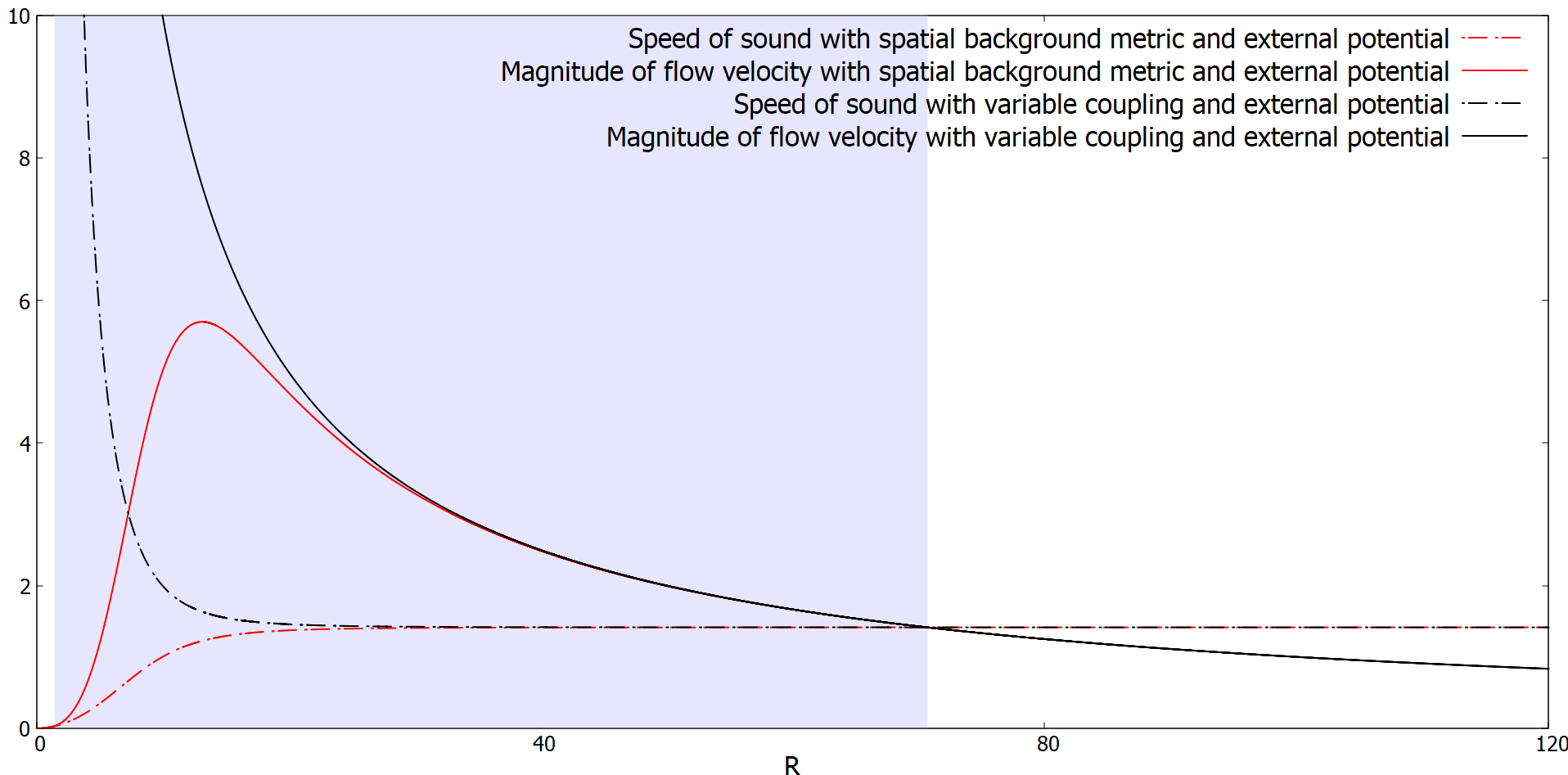}
        \caption{$|B|=50.0$ and $R_0=10.0$} \label{fig:B_50.0_r0_10.0_pot}
    \end{subfigure}
    %\vspace{5pt}
    \caption{Speed of sound (dashed line) and magnitude of flow velocity (solid line) for sample solutions with background metric (in red) and variable coupling (in black) as well as external potential in 2D with supersonic region in blue shade.}
    \label{fig:c and v with metric and pot}
\end{figure}

It is interesting to note that allowing for an external potential together with a spatial metric allows for a special class of solutions with uniform density. Indeed, setting $g_0(R)=1$ and $\rho(R)=1$ in 
\eqref{e:StationaryODEmetricA} we get
\begin{equation} \label{e:uniform}
%\begin{align}
-\frac{B^{2}}{R^{2}}-f(R) \mathcal{V}(R)   = 0
%\end{align}
\end{equation}
Therefore, such solutions exist if the potential satisfies this condition giving
\beq
\mathcal{V}\left(R \right) = -\frac{B^2}{R^2 \left(1+\left(\frac{R_0}{R}\right)^4\right)}
\eeq
for the case we are considering here where $f(R)$ is given by eq.\eqref{fR}.
With $\rho=1$, from \eqref{e:scaled_sound_funnel} and \eqref{e:scaled_vel_funnel}, the flow velocity becomes $ \Vec{\mathbf{V}}  \left(R \right) = \frac{2 B}{f\left(R \right) R} \hat{\mathbf{R}}$ and the local speed of sound becomes $C=\frac{\sqrt{2}}{\sqrt{f\left(R \right)}}$. For some sample solutions the flow speed and sound speed are shown in figure \ref{fig:c and v with metric and pot}. The advantage of this solution is that the exact location $R_h$ of the acoustic horizons is known in a closed form $R_h=\sqrt{B^2 \pm \sqrt{B^4 - {R_0}^4}}$.

\subsection{Fluctuations and acoustic wormhole metric}
It can be shown in a similar way as in the previous section that the perturbations of this background solution also look like a massless scalar field in an acoustic (space-time) metric background (from \eqref{e:acoustic metric tensor}) in the hydrodynamic limit, and the acoustic metric looks as follows: 

\begin{equation} \label{e:acoustic metric tensor bg pot}
g_{\mu \nu} \sim
\left[\begin{array}{ccc}
-\left(1-\frac{2 B^{2} R^{2}}{\left(R^{4}+R_{0}^{4}\right)}\right) & \frac{{| B |}}{R} & 0 
\\
 \frac{{| B |}}{R} & \frac{R^{4}+R_{0}^{4}}{2 R^{4}} & 0 
\\
 0 & 0 & \frac{R^{4}+R_{0}^{4}}{2 R^{2}} 
\end{array}\right]
\end{equation}

\begin{comment}
Under the coordinate transformation $\ln\left(\frac{R}{R_0} \right) = u$, \eqref{e:acoustic metric tensor bg pot} becomes,
\begin{equation} \label{e:acoustic metric tensor wormhole bg pot}
g_{\mu \nu} \sim
\left[\begin{array}{ccc}
-\left(1-\frac{B^{2}}{R_{0}^{2} \cosh \left(2 u \right)}\right) & {| B |} & 0 
\\
 {| B |} & R_{0}^{2} \cosh \! \left(2 u \right) & 0 
\\
 0 & 0 & R_{0}^{2} \cosh \! \left(2 u \right) 
\end{array}\right]
\end{equation}
which also looks like a wormhole \ref{fig:Wormhole configuration} similar to \eqref{e:acoustic metric tensor wormhole}
\end{comment}
\noindent which also looks like a wormhole (see Fig. \ref{fig:Wormhole configuration}) similar to non-uniform density configurations discussed in the previous section.

\subsection{Hawking temperature from fluctuations in the hydrodynamic limit with $R_0 \approx 0$} \label{sec:funnel_pot_R_0=0}
In the limit of $R_0 \approx 0$, this system looks like the one discussed in \cite{PhysRevA.99.023850} for a photon gas. This system has only one horizon given by $R_{h}=\sqrt{2}\lvert B\rvert$. Furthermore, $R_{0}\approx 0$ gives the flow velocity $ \Vec{\mathbf{V}}  \left(R \right) \approx \frac{2 B}{R} \hat{\mathbf{R}}$ and the local speed of sound $C\approx\sqrt{2}$ from \eqref{e:scaled_vel_funnel} and \eqref{e:scaled_sound_funnel} respectively.

\vspace{10pt}\paragraph {Bogoliubov transform:} Thinking of phase perturbations (under hydrodynamic approximation) as a massless scalar field in the acoustic metric background given by equation \eqref{e:acoustic metric tensor bg pot} with $R_0 \approx 0$, we can reproduce the system discussed in \cite{PhysRevA.99.023850}. Thus, from \cite{PhysRevA.99.023850}  we get the temperature $T_H$ of this analog black hole to be $T_H=\frac{1}{2 \pi \lvert B \rvert}$.

\vspace{10pt}\paragraph {Analytic continuation:} One more check is to analytically continue the outgoing solution in the vicinity of the horizon from outside to inside\customfootnote{The coordinates need to be smooth across the horizon without coordinate singularity which is the case with Painlev\'e-Gullstrand like coordinates of acoustic metric used.\label{analytic continuation across horizon}}{}. It can be shown that the radial dependence of the outgoing solution near the horizon has the form $(R - \sqrt{2} \lvert B \rvert)^{i \omega  \mathit{\lvert B \rvert}}$. Therefore, we get a real exponential correction inside the horizon ${\mathrm e}^{\mathit{\lvert B \rvert} \omega  \pi}$. This gives the ratio ${\mathrm e}^{-2 \mathit{\lvert B \rvert} \omega  \pi}$ (where $\omega > 0$) of the squared amplitude of the outgoing solution outside and inside the horizon, similar to the relation between Bogoliubov coefficients \cite{PhysRevA.99.023850}. This can also be thought of as a tunneling probability ($<1$) or tunneling coefficient for the analog Hawking radiation (similar to \cite{doi:10.1142/S2010194512006563} except in Painlev\'e-Gullstrand coordinates). This is consistent with the previous approach.

\vspace{10pt}\paragraph {Periodicity of Wick rotated Euclidean time:} Furthermore, if we diagonalize the acoustic metric \eqref{e:acoustic metric tensor bg pot} (with $R_0 = 0$) by combining coordinates $T$ and $R$ into a new time coordinate $T_{sch}$, the metric becomes
\beq
ds^2 = -\frac{\left(R^{2}-R_{h}^{2}\right) \mathit{dT_{sch}}^{2}}{R^{2}}+\frac{R^{2} \mathit{dR}^{2}}{2\left(R^{2}-R_{h}^{2}\right)}+\frac{R^{2}}{2} d\varphi^{2}
\eeq
where, $R_h = \sqrt{2} \lvert B \rvert$. Under the coordinate transformation $R = R_h + \frac{1}{\sqrt{2} \lvert B \rvert} \epsilon^2$, the temporal and radial parts of the metric become
\beq
-\frac{\epsilon^{2} \mathit{dT_{sch}}^{2}}{{B}^{2}}+d\epsilon^{2}
\eeq
where, through Wick rotation, if we redefine $\beta = i \frac{T_{sch}}{\lvert B \rvert}$, to avoid a conical singularity $\beta$ should have periodicity $2 \pi$ and therefore the temperature is $T_H = \frac{1}{2 \pi \lvert B \rvert}$  which gives the tunneling coefficient ${\mathrm e}^{-2 \mathit{\lvert B \rvert} \omega  \pi}$. This is consistent with the previous approaches that we discussed.

\subsection{Hawking temperature from fluctuations in the hydrodynamic limit with $R_0 \neq 0$}
Now if we go back to \eqref{e:acoustic metric tensor bg pot} and consider $R_0 \neq 0$, there are two horizons given by $R_{p,m} = \sqrt{B^2 \pm \sqrt{B^4 - {R_0}^4}}$ (in the extremal case $\lvert B\rvert = R_{0}$). Since there is no exact solution for the massless scalar filed ${\theta}_1 \left(R , \varphi , T\right)$ even in the hydrodynamic approximation, we just study the equation in the vicinity of the outer horizon.

\vspace{10pt}\paragraph{Analytic continuation:} One method is to analytically continue the solution in the vicinity of the outer horizon from the outside to the inside\footref{analytic continuation across horizon}. The radial dependence of the solution near the outer horizon $R_{p}$ is of the form $(R - R_{\mathrm{p}})^{\frac{\mathrm{i} \omega  \mathit{\lvert B \rvert}^{3}}{\sqrt{\mathit{B}^{4}-\mathit{R_0}^{4}}}}$. Therefore, we get real exponential corrections of the form ${\mathrm{e}}^{-\frac{\mathrm{\pi}  \left(R_{m}^{2}+R_{p}^{2}\right)^{\frac{3}{2}}\, \omega}{\sqrt{2}\left(R_{p}^{2}-R_{m}^{2}\right)}} = {\mathrm{e}}^{-\frac{\mathrm{\pi}  \mathit{\lvert B \rvert}^{3} \omega}{\sqrt{B^4-R_{0}^{4}}}}$ while going from inside to outside $R_{p}$. The ratio of amplitude squared outside to amplitude squared inside can be interpreted as the semi-classical probability of tunneling ($<1$) or the tunneling coefficient (similar to \cite{doi:10.1142/S2010194512006563} except in Painlev\'e-Gullstrand coordinates) giving the temperature $T_H=\frac{\sqrt{2}\left(R_{p}^{2}-R_{m}^{2}\right)}{2\mathrm{\pi}  \left(R_{m}^{2}+R_{p}^{2}\right)^{\frac{3}{2}}} = \frac{\sqrt{B^4-R_{0}^{4}}}{2\mathrm{\pi}  \mathit{\lvert B \rvert}^{3}}$.

\vspace{10pt}\paragraph{Periodicity of Wick rotated Euclidean time:} If we diagonalize the acoustic metric \eqref{e:acoustic metric tensor bg pot} by combining coordinates $T$ and $R$ into a new time coordinate $T_{sch}$, the metric becomes
\begin{align}
\lefteqn{ds^2 = -\frac{\left(R^{2}-R_{p}^{2}\right) \left(R^{2}-R_{m}^{2}\right) \mathit{dT_{sch}}^{2}}{R^{4}+R_{p}^{2} R_{m}^{2}}}\nonumber \\ &\indent\indent+\frac{\left(R^{4}+R_{p}^{2} R_{m}^{2}\right)^{2} \mathit{dR}^{2}}{2 R^{4} \left(R^{2}-R_{p}^{2}\right) \left(R^{2}-R_{m}^{2}\right)}+\frac{\left(R^{4}+R_{p}^{2} R_{m}^{2}\right) d\varphi^{2}}{2 R^{2}}
%\addtocounter{equation}{1}\tag*{\normalsize(\theequation)}
\end{align}
where, $R_p = \sqrt{B^2 + \sqrt{B^4 - {R_0}^4}}$ and $R_m = \sqrt{B^2 - \sqrt{B^4 - {R_0}^4}}$. Under the coordinate transformations near $R_p$ given by $R = R_p + \frac{\left(R_{p}^2 - R_{m}^2\right) R_{p}}{\left(R_{p}^2 + R_{m}^2\right)^{2}} \epsilon^2 = R_p + \frac{\sqrt{B^{4}-R_{0}^{4}} \left(B^{2}+\sqrt{B^{4}-R_{0}^{4}}\right)}{2 \lvert B\rvert^{4}} \epsilon^2$, the temporal and radial parts of the metric become
\beq
\text{At }R_{p} \implies -\frac{2 \left(-R_{m}^{2}+R_{p}^{2}\right)^{2}}{\left(R_{m}^{2}+R_{p}^{2}\right)^{3}}\epsilon^{2} \mathit{dT_{sch}}^{2}+d\epsilon^{2} = -\frac{B^{4}-R_{0}^{4}}{B^{6}}\epsilon^{2} \mathit{dT_{sch}}^{2}+d\epsilon^{2}
\eeq
where, through Wick rotated Euclidean time and its periodicity of $2 \mathrm{\pi}$, we get the temperature at outer Black hole horizon
\beq
T_H=\frac{\sqrt{2}\, \left(R_{p}^{2}-R_{m}^{2}\right)}{2 \mathrm{\pi} \left(R_{m}^{2}+R_{p}^{2}\right)^\frac{3}{2}} = \frac{\sqrt{B^{4}-R_{0}^{4}}}{2\mathrm{\pi}{\lvert B \rvert}^{3}}
\eeq
This gives tunneling coefficient across the outer horizon
\beq
{\mathrm{e}}^{-\frac{2 \mathrm{\pi} \omega \left(R_{m}^{2}+R_{p}^{2}\right)^\frac{3}{2}}{\sqrt{2}\, \left(R_{p}^{2}-R_{m}^{2}\right)}} = {\mathrm{e}}^{-\frac{2\mathrm{\pi}\omega {\lvert B \rvert}^{3}}{\sqrt{B^{4}-R_{0}^{4}}}}
\eeq
This is consistent with the analytic continuation across the horizon.

\section{Radially varying coupling and external potential}
In a laboratory setting, a non-trivial background metric can be accomplished by restricting the fluid to move on a curved surface rather than a flat one. Although this might be possible, we point out in this section that the solution corresponding to the spatial funnel metric in section \ref{sec:wormhole BVP} can be reinterpreted as a solution in a spatial flat metric by using the correspondence principle discussed in section \ref{sec:general}. Instead of $f(R)$, $\mathcal{V}(R)$ and $g_0=1$ we can use a flat metric with potential $U(R)$ and coupling $g_0(R)$ such that:
\begin{align}
U(R) - 1 & = \left(\mathcal{V}(R) - 1 \right) f(R)  \label{e:H}\\
g_0(R) & = f(R) \label{e:I}
\end{align}
where $f(R) = 1 + \left(\frac{R_0}{R} \right)^4$. Both of these configurations (on the left and the right above) have the identical stationary solutions. Notice that, with the funnel metric, the region near $R=0$ was interpreted as another asymptotic region as is evident from the symmetry between large and small $R$. The incoming flux from one region goes out through the other region. On the other hand, in the corresponding flat metric with variable coupling case, $R=0$ represents the origin of the plane and the fluid appears to have nowhere to go. By carefully looking at the GP equation one can see that the solution corresponds to a GP equation with a sink at the center that removes the atoms, which can be simulated by an imaginary Dirac delta-function potential at $R=0$. This is also reflected in the fact that the scaled local speed of sound is now $C = \sqrt{2 f(R)} \rho (R)$ and the flow velocity is $\Vec{\mathbf{V}}  \left(R \right) = \frac{2 B}{R (\rho\left(R \right))^2} \hat{\mathbf{R}}$. 
They are singular when $R\rightarrow 0$ (see Fig. \ref{fig:c and v with metric}) despite the fact that the amplitude/density of the stationary solution itself is well behaved (see Fig. \ref{fig:sample solution with metric}). This breaks the symmetry about $R_{0}$ which is also evident from the equations that govern fluctuations, as we see in the next subsection.

\subsection{Fluctuations}
Introducing perturbation in the stationary solution as done earlier \eqref{e:den phase pert} we find
\begin{equation} \label{e:reinterpreted_funnel_density_pert}
%\begin{align}
\left({\partial}_T - \frac{2 \lvert B\rvert}{R (\rho_{0} \left(R \right))^2} {\partial}_R \right) N_1 \left(R , \varphi , T\right) = - \frac{1}{{{\rho}_0 (R)}^2} {\nabla}_R \cdot ({{\rho}_0 (R)}^2 {\nabla}_R {\theta}_1 \left(R , \varphi , T\right))
%\end{align}
\end{equation}

\begin{align} \label{e:reinterpreted_funnel_phase_pert}
\centering
%\begin{align}
\left({\partial}_T - \frac{2 \lvert B\rvert}{R (\rho_{0} \left(R \right))^2} {\partial}_R \right) {\theta}_1 \left(R , \varphi , T\right) & = \frac{1}{{{\rho}_0 (R)}^2} {\nabla}_R \cdot ({{\rho}_0 (R)}^2 {\nabla}_R N_1 \left(R , \varphi , T\right)) \nonumber \\ & - 2 f(R) {{\rho}_0 (R)}^2 N_1 \left(R , \varphi , T\right)
%\end{align}
\end{align}
where, $N_1 \left(R , \varphi , T\right) = \frac{n_{1} \! \left(R , \varphi , T\right)}{2 n_{0} \! \left(R \right)} = \frac{n_{1} \! \left(R , \varphi , T\right)}{2 \left(\rho_{0} \! \left(R \right) \right)^2}$. We notice that  \eqref{e:reinterpreted_funnel_density_pert} and \eqref{e:reinterpreted_funnel_phase_pert} clearly differ from \eqref{e:funnel_density_pert} and \eqref{e:funnel_phase_pert}. In addition, equations \eqref{e:reinterpreted_funnel_density_pert} and \eqref{e:reinterpreted_funnel_phase_pert}  lack the symmetry about $R_{0}$ that \eqref{e:funnel_density_pert} and \eqref{e:funnel_phase_pert} enjoy. Now we proceed to analyze the solutions. 

Under the hydrodynamic approximation ${\nabla}_R \cdot ({{\rho}_0 (R)}^2 {\nabla}_R N_1 \left(R , \varphi , T\right)) \approx 0$, equations \eqref{e:reinterpreted_funnel_density_pert} and \eqref{e:reinterpreted_funnel_phase_pert} can be combined, and the result looks like a massless scalar field $\theta_{1} \left(R , \varphi , T\right)$ in an acoustic (space-time) metric background (see \cite{PhysRevA.63.023611}). The acoustic metric here looks like the Schwarzschild metric in Painlev\'e-Gullstrand coordinates (see \cite{HamiltonAndrewJ.S.2008Trmo}).

\begin{align} \label{e:reinterpreted acoustic metric tensor}
g_{\mu \nu} & \sim \rho_{0} \! \left(R \right)^{4}
\left[\begin{array}{ccc}
-\left(1-\frac{2 B^{2} R^{2}}{\rho_{0} \! \left(R \right)^{6} \left(R^{4}+R_{0}^{4}\right)}\right) & \frac{{| B | R^3}}{\left(R^{4}+R_{0}^{4}\right) \rho_{0} \left(R \right)^{4}} & 0 
\\
 \frac{{| B | R^3}}{\left(R^{4}+R_{0}^{4}\right) \rho_{0} \left(R \right)^{4}} & \frac{R^{4}}{2 \left(R^{4}+R_{0}^{4}\right) \rho_{0} \left(R \right)^{2}} & 0 
\\
 0 & 0 & \frac{R^{6}}{2 \left(R^{4}+R_{0}^{4}\right) \rho_{0} \left(R \right)^{2}} 
\end{array}\right]
\end{align}
This acoustic metric tensor clearly lacks the symmetry about $R_{0}$ and it no longer corresponds to a (one-way) wormhole-like configuration because it lacks a second asymptotic region at $R=0$.

\section{Uniform density solution with radially varying coupling and external potentials} \label{sec:reinterpreted_funnel_w_external_pot_additional_external_pot}
From equation \eqref{e:StationaryODEmetricA} we find that there are uniform density solutions $\rho=1$ with no external metric if the condition 
\begin{equation} \label{e:uniform2}
%\begin{align}
-\frac{B^{2}}{R^{2}}+ \left[1-\mathcal{V}(R)-g_0(R) \right]  = 0
%\end{align}
\end{equation}
is satisfied. Using the correspondence principle of section \ref{sec:general} we find this case can be related to a non-trivial metric $f(R)$, potential $\tilde{\mathcal{V}}(R)$ and constant coupling if:
\beqa
 (1-\mathcal{V}(R))  &=&  f(R) (1-\tilde{\mathcal{V}}(R))  \\
 g_{0}(R)  &=& f(R) 
\eeqa 
 Therefore, setting 
\beq
 g_0(R) = 1+\left(\frac{R_0}{R}\right)^4
\eeq
and the potential according to \eqref{e:uniform2} we find uniform density solutions with a non-trivial acoustic metric for the fluctuations as discussed in the next subsection. Furthermore, the local speed of sound and the flow velocity become $C = \sqrt{2 f(R)}$ and $\Vec{\mathbf{V}}  \left(R \right) = \frac{2 B}{R} \hat{\mathbf{R}}$, respectively, which as shown in Figure \ref{fig:c and v with metric and pot} are clearly singular\customfootnote{Inner horizon is excluded from the plot because it is way outside the bounds of the plot. This is because sound and flow speeds are singular as discussed.} despite the $\rho (R)$ of the stationary background solution being $1$ everywhere. These are perhaps the solutions that are easier to reproduce in an experimental setting. 

\subsection{Fluctuations}
It can be shown in a similar way as in the previous section that the perturbations in this background solution also look like a massless scalar field in an acoustic (space-time) metric background (from \eqref{e:reinterpreted acoustic metric tensor}) in the hydrodynamic limit and the acoustic metric looks as follows:
\begin{align} \label{e:reinterpreted acoustic metric tensor additional potential}
g_{\mu \nu} & \sim
\left[\begin{array}{ccc}
-\left(1-\frac{2 B^{2} R^{2}}{\left(R^{4}+R_{0}^{4}\right)}\right) & \frac{{| B | R^3}}{\left(R^{4}+R_{0}^{4}\right)} & 0 
\\
 \frac{{| B | R^3}}{\left(R^{4}+R_{0}^{4}\right)} & \frac{R^{4}}{2 \left(R^{4}+R_{0}^{4}\right)} & 0 
\\
 0 & 0 & \frac{R^{6}}{2 \left(R^{4}+R_{0}^{4}\right)} 
\end{array}\right]
\end{align}
This metric tensor also clearly lacks the symmetry about $R_{0}$ and it no longer corresponds to a wormhole because it lacks a second asymptotic region.

\subsection{Hawking temperature from fluctuations in the hydrodynamic limit with $R_0 \approx 0$}
In the limit of $R_0 \approx 0$, metric tensors \eqref{e:reinterpreted acoustic metric tensor additional potential} and \eqref{e:acoustic metric tensor bg pot} look exactly the same and therefore with $R_0 \approx 0$ this system also looks like the one discussed in \cite{PhysRevA.99.023850} for the photon gas and therefore the solutions of the perturbation equations exhibit the same behavior as discussed for the uniform density solution with spatial background funnel metric and external potential. Therefore, the temperature is $T_H=\frac{1}{2 \pi \lvert B \rvert}$  which gives the tunneling coefficient ${\mathrm e}^{-2 \mathit{\lvert B \rvert} \omega  \pi}$.

\subsection{Hawking temperature from fluctuations in the hydrodynamic limit with $R_0 \neq 0$}
Now, if we go back to \eqref{e:reinterpreted acoustic metric tensor additional potential} and consider $R_0 \neq 0$, there are two horizons given by $R_{p,m} = \sqrt{B^2 \pm \sqrt{B^4 - {R_0}^4}}$ (in the extremal case $\lvert B\rvert = R_{0}$) but again there is no exact solution for the massless scalar ${\theta}_1 \left(R , \varphi , T\right)$ even in the hydrodynamic approximation. However, we can still analyze the behavior near the outer horizon as follows.

\vspace{10pt}\paragraph{Analytic continuation:} One can analytically continue the solution in the vicinity of the outer horizon from the outside to the inside\footref{analytic continuation across horizon}. In this case, the solution near the outer horizon $R_{p}$ has the radial dependence of the form $(R - R_{\mathrm{p}})^{\frac{\mathrm{i} \omega  \mathit{\lvert B \rvert} \left(\mathit{B}^{2}+\sqrt{\mathit{B}^{4}-\mathit{R_0}^{4}} \right)}{2 \sqrt{\mathit{B}^{4}-\mathit{R_0}^{4}}}}$. Therefore, we get real exponential corrections while going from outside to inside $R_{p}$. The ratio of the modulus square of the solution outside and inside the horizon can be interpreted as a tunneling coefficient (similar to \cite{doi:10.1142/S2010194512006563} except in Painlev\'e-Gullstrand coordinates) or the probability of tunneling ($<1$). This indicates the temperature $T_H=\frac{\sqrt{2}\left(R_{p}^{2}-R_{m}^{2}\right)}{2 \mathrm{\pi}  \sqrt{R_{m}^{2}+R_{p}^{2}} R_{p}^{2} } = \frac{2\sqrt{B^4-R_{0}^{4}}}{2\mathrm{\pi}  \mathit{\lvert B \rvert} \left(B^2+\sqrt{B^4-R_{0}^{4}}\right)}$.

\vspace{10pt}\paragraph{Periodicity of Wick rotated Euclidean time:} If we diagonalize the acoustic metric \eqref{e:reinterpreted acoustic metric tensor additional potential} by combining coordinates $T$ and $R$ into a new time coordinate $T_{sch}$, the metric becomes
\begin{align}
\lefteqn{ds^2 = -\frac{\left(R^{2}-R_{p}^{2}\right) \left(R^{2}-R_{m}^{2}\right) \mathit{dT_{sch}}^{2}}{R^{4}+R_{p}^{2} R_{m}^{2}}}\nonumber\\ &\indent\indent+\frac{R^{4} \mathit{dR}^{2}}{2 \left(R^{2}-R_{p}^{2}\right) \left(R^{2}-R_{m}^{2}\right)}+\frac{R^{6}}{2 R^{4}+2 R_{p}^{2} R_{m}^{2}} d\varphi^{2}
%\addtocounter{equation}{1}\tag*{\normalsize(\theequation)}
\end{align}
where, $R_p = \sqrt{B^2 + \sqrt{B^4 - {R_0}^4}}$ and $R_m = \sqrt{B^2 - \sqrt{B^4 - {R_0}^4}}$. Under the coordinate transformations near $R_p$ given by $R = R_p + \frac{R{p}^2 - R_{m}^2}{R_{p}^{3}} \epsilon^2 = R_p + \frac{2 \sqrt{B^{4}-R_{0}^{4}}}{\left(B^{2}+\sqrt{B^{4}-R_{0}^{4}}\right)^{\frac{3}{2}}} \epsilon^2$, the temporal and radial parts of the metric become
\begin{align}
\text{At }R_{p} \implies& -\frac{2 \left(-R_{m}^{2}+R_{p}^{2}\right)^{2}}{R_{p}^{4} \left(R_{m}^{2}+R_{p}^{2}\right)}\epsilon^{2} \mathit{dT_{sch}}^{2}+d\epsilon^{2}\nonumber\\ =& -\frac{4 B^{4}-4 R_{0}^{4}}{\left(B^{2}+\sqrt{B^{4}-R_{0}^{4}}\right)^{2} B^{2}}\epsilon^{2} \mathit{dT_{sch}}^{2}+d\epsilon^{2}
%\addtocounter{equation}{1}\tag*{\normalsize(\theequation)}
\end{align}
where, through Wick-rotated Euclidean time (see the process explained in \cite{osti_6637168}) and its periodicity of $2 \mathrm{\pi}$ we get the temperature at the outer Black hole horizon
\beq
T_H=\frac{\sqrt{2}\, \left(R_{p}^{2}-R_{m}^{2}\right)}{2 \mathrm{\pi} R_{p}^{2} \sqrt{R_{m}^{2}+R_{p}^{2}}} = \frac{2 \sqrt{B^{4}-R_{0}^{4}}}{2 \mathrm{\pi} \left(B^{2}+\sqrt{B^{4}-R_{0}^{4}}\right) \lvert B \rvert}
\eeq
This gives a tunneling coefficient
\beq
{\mathrm{e}}^{-\frac{2 \mathrm{\pi} \omega R_{p}^{2} \sqrt{R_{m}^{2}+R_{p}^{2}}}{\sqrt{2}\, \left(R_{p}^{2}-R_{m}^{2}\right)}} = {\mathrm{e}}^{-\frac{2 \mathrm{\pi} \omega \left(B^{2}+\sqrt{B^{4}-R_{0}^{4}}\right) \lvert B \rvert}{2 \sqrt{B^{4}-R_{0}^{4}}}} 
\eeq
consistent with the analytic continuation across the horizon.

\section{Conclusions}
We obtained non-singular stationary solutions of the Gross-Pitaevskii equation by putting a spatial funnel-like metric describing a space with two asymptotic regions. The fluid moves radially inwards 
in one region and comes out in the other one. Therefore the fluid does not accumulate and no singularity is produced. The spatial funnel-metric is particularly interesting because it resembles a (one-way) wormhole for phonons in the hydrodynamic limit. We also obtained non-singular stationary solutions of the Gross-Pitaevskii equation using position dependent coupling and potential. We expect this latter approach to be more feasible in actual experiments. In this approach, we do not see a wormhole-like behavior, but we still do get acoustic black/white hole configurations. Perhaps the most interesting configurations from a practical point of view  
are ones with uniform density that appear when the funnel metric or the position dependent coupling and potential are related to each other in a particular way. 

In all the configurations discussed, the approximate local speed of sound and the magnitude of flow velocity cross. That means we do indeed find acoustic black/white hole and wormhole configurations. We also see that, in the hydrodynamic limit, the phase fluctuations around the stationary solutions behave as a massless scalar (phonon) field in an acoustic metric background. In particular, the fluctuations around the uniform density solutions are easy to analyze, leading to specific values for the black-hole temperature.  

\section{Acknowledgments}

We are grateful to the DOE for support under the Fermilab Quantum Consortium. We are grateful to Dr. Sergei Khlebnikov and Dr. Chen-Lung Hung for various comments and discussions. We also wish to acknowledge valuable computational resources provided by RCAC (Rosen Center for Advanced Computing) at Purdue University.

%\nocite{*}
\printbibliography

\appendix
\numberwithin{equation}{section}

\newpage
\section{Acoustic wormhole double-funnel} \label{Appendix A}

In section \ref{sec:wormhole BVP} we consider a 2D quantum fluid in a conformally flat metric given by 
\beq\label{metricfR}
 ds^2 = f(R)\, (dR^2 + R^2 d\theta^2), \ \ \ \ f(R) = 1+\frac{R_0^4}{R^4}
\eeq
To potentially recreate such metric in a laboratory setting we should force the fluid to move on a curved surface with the shape of a double funnel as in fig. \ref{fig:Wormhole configuration}. To find the exact required shape we consider, in cylindrical coordinates a surface parameterized by $(R,\theta)$ of the form:
\beq
 r=r(R), \ \ \theta=\theta, \ \ z=z(R) 
\eeq
such that the induced metric on the surface
\beqa
 ds^2 &=& dr^2 + r^2 d\theta^2 + dz^2 \\
      &=& \left(\frac{\partial r(R)}{\partial R}\right)^2 dR^2 + r(R)^2 d\theta^2 +   \left(\frac{\partial z(R)}{\partial R}\right)^2 dR^2 
\eeqa
agrees with \eqref{metricfR}. From this we get the following equations
\beqa \label{e:z(R)_ODE}
\frac{d z \! \left(\mathit{R} \right)}{d \mathit{R}} &= \frac{2 \,R_{0}^{2}}{\sqrt{\mathit{R}^{4}+R_{0}^{4}}} \\
r(R) &= \sqrt{f(R)} R
\eeqa
The function $r(R)$ is determined and the equation for $z(R)$ can be integrated in terms of elliptic functions. Making a parametric plot of $r(R)$ and $z(R)$, we get the double-funnel geometry shown (with scaled out $R_0$) in Figure \ref{fig:Wormhole configuration}. This shows that the metric is flat at $R\to\infty\implies r(R)\to\infty$, $z(R)>0$ and $R\to 0\implies r(R)\to\infty$, $z(R)<0$. The span of the coordinate $z$ is $\Delta z = z(R=+\infty)-z(R=0) = \frac{R_0}{2}\frac{\left(\Gamma(\frac{1}{4})\right)^{2}}{\Gamma(\frac{1}{2})}$

Another useful radial coordinate that respects the symmetry is $\xi = \frac{R}{R_0} - \frac{R_0}{R}$  with $-\infty<\xi<\infty$.

\end{document}